\documentclass{article}

\usepackage{amsmath,amsthm,amssymb}
\usepackage{graphics,graphicx}
\usepackage{epsfig}
\usepackage{cite}

\newcommand{\be}[1]{\begin{equation} #1 \end{equation}}

\newcommand{\paren}[1]{\left( #1 \right)}

\newcommand{\dis}{\displaystyle}
\newcommand{\goodspace}{\hspace{0.6cm}}

\newcommand{\onehalf}{\frac{1}{2}}

\def\eg{{\it e.g.\ }}
\def\etal{{\it et al.\ }}
\def\ie{{\it i.e.\ }}

\begin{document}

{\hfill USITP 07-02} \\[-5mm]

{\hfill 14 November 2007}

\begin{center}
\textbf{\Large On Geometro-thermodynamics of Dilaton Black Holes}\\
\vspace{1cm}
\normalsize {\bf Jan E. \AA man}\footnote{ja@physto.se}, {\bf Narit Pidokrajt}\footnote{narit@physto.se} \\[.2cm]
\emph{Fysikum, AlbaNova \\
Stockholm University\\
SE-10691 Stockholm\\
Sweden} \\[.4cm]

\normalsize{\bf John Ward}\footnote{jwa@uvic.ca}\\[.2cm]
\emph{Department of Physics and Astronomy \\ University of Victoria\\ Victoria, BC, V8P 1A1, Canada}
\end{center}

\begin{abstract}
In this talk we present the latest results from our ongoing project on geometro-thermodynamics (also known as information geometry of thermodynamics or Ruppeiner geometry) of dilaton BHs in 4D in both Einstein and string frames and a dyonic dilaton BH and at the end we report very briefly results from this approach to the 2D dilaton BHs. 
\end{abstract}

The thermodynamic geometry, also known as Ruppeiner geometry (Ruppeiner \cite{ruppeiner1, ruppeiner2}), of various BH families has been studied over the past few years (see, \eg {\AA}man \etal \cite{ourpaper1}-\cite{ourpaper4}, Arcioni \etal \cite{arcioni}, Sarkar \etal \cite{sarkar}, Shen \etal \cite{shen}, Mirza \etal \cite{mirza}, Ruppeiner \cite{ruppeiner} and Quevedo \cite{quevedo}). Our results so far have been physically suggestive, particularly in the Myers-Perry Kerr BH case where the curvature singularities signal thermodynamic instability of the BH. The geometrical patterns are given by the curvature of the Ruppeiner metric\footnote{This metric is conformal to the so-called Weinhold (Weinhold \cite{weinhold}) metric via $g^W_{ij} = T g^R_{ij}$ where $T$ is thermodynamic temperature of the system of interest.} defined as the Hessian of the entropy on the phase space of the thermodynamic system 
\be{
g^R_{ij} = - \partial_i \partial_j S(M, N^a),
}
where $M$ denotes mass (internal energy) and $N^a$ are other parameters such as charge and spin. The minus sign
arises because entropy is a concave function. Interpretations of the geometries associated with the metric are discussed in Ruppeiner \cite{ruppeiner2} and references therein.  Even though most interesting Ruppeiner metrics that we encounter have curvature singularities which might be interpretable, there are some known flat Ruppeiner metrics that shed light on the understanding of thermodynamic geometries as a whole \ie the structure of the entropy function (or mass) that gives a flat Ruppeiner geometry. In  {\AA}man  \etal \cite{ourpaper2} we proved a  flatness theorem which states that Riemann curvature tensor constructed out of the negative of the Hessian of the entropy of the form 
\be{
S = M^k f(Q/M)
}
will vanish, where $f$ is an arbitrary analytic function and $k \neq 1$. The latter condition is necessary in order for the metric to be nondegenerate. This theorem has proven useful in our work on the dilaton BHs as it allows us to see the local geometry already by glancing at the entropy function.

In this short article we will focus on the ongoing work on the application of the geometrical approach to the thermodynamics of the dilaton BHs. The 4D dilaton BH\footnote{There are many references on this BH but we refer to Garfinkle \etal \cite{dilaton1}, Frolov \etal \cite{book} and Casadio \etal \cite{stringframe} for the dilaton BH in string frame.} action can be obtained from the low energy limit of the string theory\footnote{In fact $a=1$ for the dilaton gravity reduced from string theory but for our work we use arbitrary $a$ which determines the gravitational coupling strength.} by dropping certain terms except for the metric $g_{\mu\nu}$, a dilaton $\phi$ and a Maxwell field $F_{\mu\nu}$. The action thus obtained is 
\be{
S_{st} = \int d^4 x \sqrt{-g} e^{-2a \phi} \left(R + 4(\nabla \phi)^2-F^2 \right).
} 
By making a conformal transformation  $g_{\mu\nu}^{\textrm{\tiny string}} = e^{2a\phi} g_{\mu\nu}^{\textrm{\tiny Einstein}}$ we obtain the dilaton gravity action in Einstein frame 
\be{
S_E = \int d^4 x \sqrt{-g} \paren{ R - 2(\nabla \phi)^2 - e^{-2a \phi}F^2}.
}
In the absence of the Maxwell field the action reduces to the standard Einstein theory with a massless scalar field as the matter.  The BH solution of the two actions is given by the standard spherically symmetric metric 
\be{
ds^2 = -N dt^2 + \frac{1}{N}dr^2 + P^2(r) d\Omega^2_2
}
where $d\Omega^2_2$ is the usual 2-sphere metric and 
\be{
N = \left(1-\frac{r_{+}}{r}\right)\left(1-\frac{r_{-}}{r}\right)^{\frac{(1+a)^2}{1+a^2}}, \goodspace 
P = r\left(1-\frac{r_{-}}{r}\right)^{\epsilon(a)}.
}
The function $\epsilon(a)$ is different for both frames, \ie 
\be{
\epsilon(a)_{Einstein} = \frac{a^2}{1+a^2},\goodspace \epsilon(a)_{string} = \frac{a(1+a)}{1+a^2}.
}
The horizon is located at $r = r_+$ with the singularity of the solution at $r = r_-$ for $a \neq 0$. The entropy for both BHs are given by
\be{
\small
S = M^2\paren{ 1 + \sqrt{1 - (1-a^2)\frac{Q^2}{M^2}}}^2\paren{1 -  \frac{\dis (1 + a^2)Q^2}{M^2\paren{\dis 1 + \sqrt{1 - (1-a^2)\frac{Q^2}{M^2}}}^2}}^{\epsilon(a)}.
\label{the-entropy}
}
The dyonic dilaton BH is slightly more complicated in that there is another $U(1)$ gauge\footnote{We denote the magnetic charge by $P$ in contrast to the electric charge $Q$.} field coupled to the action which takes the form
\be{
S = \int d^4 x \sqrt{-g} \paren{ R - 2(\nabla \phi)^2 - e^{-2a \phi}(F^2 + G^2)}.
}
We only consider such BHs in Einstein frame. The entropy of the dyonic dilaton BH will have two charges hence not compatible with the flatness theorem. \\

%-----------------------------------------------------------
\begin{figure}[t]
\centering
\includegraphics[width=9cm]{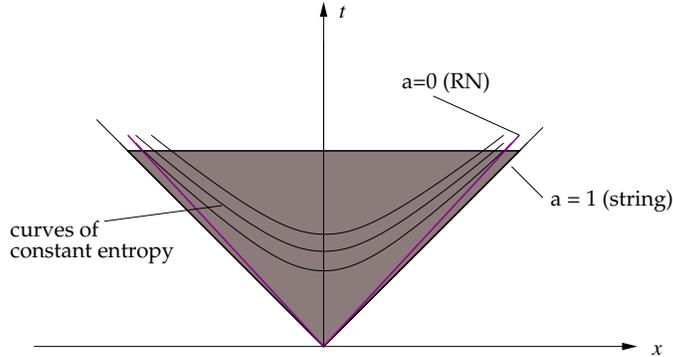}
\caption{\sl The state space of the dilaton BH in Einstein frame with a coupling constant $a \neq 0$ is a 
wedge on the null cone. For the dilaton BH in string frame the wedge fills the null cone for $a=0$ and 1.}
\end{figure}
%-----------------------------------------------------------

The Ruppeiner metric for the 4D dilaton BH in both frames is flat for any values of $a$ as can be seen using our flatness theorem. In order to illustrate the state space of the flat Ruppeiner metric for both BHs we employ Eq. (11) which is a diagonal metric derived from the entropy in Eq. (\ref{the-entropy}) given by  
\be{
ds^2_R = \frac{-1}{2S}dS^2 + g(u)du^2
\label{diagonal}
}
where $u = Q/M$ and $g(u) = \onehalf \frac{f'^2}{f^2} - \frac{f''}{f}$. The function $f = f(u)$ and its derivatives are basically the parentheses in Eq. (\ref{the-entropy}) and for an arbitrary $a$ it is very complicated. However a power-series expansion of $g(u)$ allows us to determine its range thereby giving us information on the opening of the wedge when we transform the metric (\ref{diagonal}) into a flat Minskowski via Rindler coordinates ($ds^2 = -d\tau^2 + \tau^2 d\sigma^2$). In our previous works, the wedge fills in the null cone for the Myers-Kerr BH when $D>5$\footnote{We have also found that the curvature scalar of this BH diverges where there is a thermodynamic instability which occurs in the extremal limit. The null cone represents the extremal limit where temperature vanishes.}. In this work we find that the dilaton BH in the Einstein frame the wedge fills the null cone for any value of $a \neq 0$ whereas the wedge in string frame fills the null cone of any value of $a \neq 0$ and 1 . The Ruppeiner metric and its conformal metric for the dyonic dilaton BH is nonflat with curvature singularity in the extremal limit. This was done using the CLASSI program ({\AA}man \cite{classi}). In the limit of $Q^2 = P^2$ the geometry becomes flat as expected.

We also studied the Ruppeiner geometry of the 2D dilaton BH solutions which contain the Reissner-Nordstr\"om-like BH solution and BTZ (for a comprehensive review, see Grumiller \cite{daniel, daniel1}). Results up to date are enlightening and consistent with results in higher dimensions and our flatness theorem, as summarized in {\AA}man \etal \cite{ourpaper4}. However they await further interpretations.  
 
We conclude this article by noting that the geometro-thermodynamics of BHs provides an alternative 
 route to obtain insight into thermodynamics through Riemannian geometry. Certain features of the geometry yield results that are consistent with those in the literature whereas some other features can potentially give rise to a deeper understanding of BH thermodynamics and perhaps the underlying statistical mechanics of BHs. Despite the fact that only few results so far are physically suggestive, the geometrical patterns we have observed may play an important role in the future, when quantum gravity is better understood. 

\section*{Acknowledgments}
N.P. is supported by Doktorandtj\"anst of Stockholm University and is grateful the Johan S\"oderbergs stiftelsen for a scholarship. He thanks the LOC of ERE2007 for their kind support, Alba G. Pedemonte and Tolga Birkandan for stimulating discussions. Ingemar Bengtsson and Daniel Grumiller are acknowledged for useful suggestions.


\begin{thebibliography}{99}
\bibitem[1979]{ruppeiner1} Ruppeiner, G. 1979, Phys. Rev. A., 20, 1608.
\bibitem[1995]{ruppeiner2} Ruppeiner, G. 1995, Rev. Mod. Phys., 67, 605.
\bibitem[1975]{weinhold} Weinhold, F. 1975, Chem. Phys., 63, 2479.
\bibitem[2003]{ourpaper1} \AA man, J. E. , Bengtsson, I. \& Pidokrajt, N. 2003, Gen. Rel. Grav., 35, 1733.
\bibitem[2006]{ourpaper2} \AA man, J. E. \& Pidokrajt, N. 2006, Phys. Rev. D., 73, 024017.
\bibitem[2006]{ourpaper3} \AA man, J. E. , Bengtsson, I. \& Pidokrajt, N., Gen. Rel. Grav., 38, 1305.
\bibitem[2007]{ourpaper4} \AA man, J. E. , Bedford, J., Grumiller, D., Pidokrajt, N. \& Ward, J. 2007, J. Phys.: Conf. Ser., 66, 012007.
\bibitem[2005]{arcioni} Arcioni, G., Lozano-Tellechea, E. 2005, Phys. Rev. D., 72, 104021.
%\bibitem[2007]{jan2007} \AA man, J. E. , Bengtsson, I. \& Pidokrajt, N. 2007, submitted to The proceedings volume of the "Encuentros Relativistas Espa{\~n}oles - Spanish Relativity Meeting ERE07"
\bibitem[2006]{sarkar} Sarkar, T., Sengupta, G. \& Tiwari, B.N. 2006, JHEP, 0611, 015.
\bibitem[2006]{shen} Shen, J., Cai, R-G., Wang, B. \& Su, R-K. 2006, Int. J. Mod. Phys., A22, 11-27.
\bibitem[2007]{mirza} Mirza, B. \& Zamani-Nasab, M. 2007, JHEP, 06, 059.
\bibitem[2007]{quevedo} Quevedo, H. 2007, arXiv:0704.3102 
\bibitem[2007]{ruppeiner} Ruppeiner, G. 2007,  Phys. Rev. D., 75, 024037.
\bibitem[1991]{dilaton1}  Garfinkle, D., Horowitz, G.T. \& Strominger, A. 1991, Phys. Rev. D., 43, 3140.
\bibitem[1998]{book} Frolov, V.P. \& Novikov, I.D. 1998, {\it Black Hole Physics--Basic Concepts and New Developments}, Kluwer Academic Publishers, the Netherlands.
\bibitem[1999]{stringframe} Casadio, R. \& Harms, B. 1999, 	Mod. Phys. Lett. A, 14, 1089.
\bibitem[2002]{classi} \AA man, J. E. 2002, {\it Manual for CLASSI: Classification Programs for Geometries in General Relativity}, Department of Physics, Stockholm University, Technical Report, Provisional Edition, Distributed with the sources for SHEEP and CLASSI.
\bibitem[2002]{daniel} Grumiller D., Kummer W. \& Vassilevich, D. V. 2002, Phys. Rept., 369, 327. 
\bibitem[2006]{daniel1} Grumiller D. \& Meyer, R. 2006, Turk. J. Phys., 30, 349.
\end{thebibliography}
\end{document}